\documentclass[superscriptaddress,amsmath,amssymb,nofootinbib,eqsecnum,a4paper,secnumarabic,11pt]{revtex4}         
\usepackage[pdftex]{graphicx}
\usepackage{hyperref}     
\newcommand{\mDM}{m_{\text{DM}}}

\usepackage{titlesec}     
\newcommand*{\justifyheading}{\raggedright}
\titleformat{\chapter}[display]
  {\normalfont\huge\bfseries\justifyheading}{\chaptertitlename\ \thechapter}
  {20pt}{\Huge}
\titleformat{\section}
  {\normalfont\Large\bfseries\justifyheading}{\thesection}{1em}{}
\titleformat{\subsection}
  {\normalfont\large\bfseries\justifyheading}{\thesubsection}{1em}{}
\titleformat{\subsubsection}
  {\normalfont\bfseries\justifyheading}{\thesubsection}{1em}{}
\usepackage[english]{babel}

\begin{document}

\title{ Effect of CP violation in the singlet-doublet dark matter model }
\author{Tomohiro Abe}
\affiliation{
  Institute for Advanced Research, Nagoya University,
  Furo-cho Chikusa-ku, Nagoya, Aichi, 464-8602 Japan
}
\affiliation{
  Kobayashi-Maskawa Institute for the Origin of Particles and the
  Universe, Nagoya University,
  Furo-cho Chikusa-ku, Nagoya, Aichi, 464-8602 Japan
}

\begin{abstract}
We revisit the singlet-doublet dark matter model with a special emphasis on the effect of CP violation on the dark matter phenomenology.
The CP violation in the dark sector induces a pseudoscalar interaction of a fermionic dark matter candidate with the SM Higgs boson. 
The pseudoscalar interaction helps the dark matter candidate evade the strong constraints from the dark matter direct detection experiments. 
We show that the model can explain the measured value of the dark matter density 
even if dark matter direct detection experiments do not observe any signal. 
We also show that the electron electric dipole moment is an important complement to the direct detection for testing this model. 
Its value is smaller than the current upper bound but within the reach of future experiments.
\end{abstract}


\maketitle

\section{Introduction}

Dark matter (DM) is a leading candidate for physics beyond the standard model of particle physics.
Models based on the WIMP paradigm are popular and have been widely studied. 
On the other hand, 
the recent dark matter direct detection experiments~\cite{1608.07648, 1611.06553} give severe constraints on the models.

It is possible to evade the constraints from the direct detection experiments 
if a DM candidate is a fermion and interacts with the standard model (SM) sector through pseudoscalar interactions~\cite{Escudero:2016gzx, Escudero:2016kpw}.
In that case, the cross section for the direct detection is suppressed by the velocity of the DM and thus is negligible.
On the other hand, the same interactions are relevant to DM annihilation processes
and we can obtain the DM thermal relic abundance that matches the measured value of the DM density.

There are two simple ways to introduce the pseudoscalar interactions.
One way is to add CP-odd scalar mediators that couple both to a DM candidate and the SM particles~\cite{1401.6458, 1403.3401, 1408.4929, 1501.07275,
1502.06000,1506.03116,1509.01110,1609.09079,1611.04593,1612.06462,1701.04131,1701.07427}. 
The other way is to introduce CP-violation into the dark sector. 
In the latter case, the SM Higgs boson can be a mediator, and we do not need CP-odd scalar mediators.
We focus on the latter possibility.

We consider the singlet-doublet model~\cite{hep-ph/0510064,0705.4493,0706.0918}.
This model is one of the minimal setups in simplified dark matter models with a fermionic dark matter candidate 
and has been widely studied~\cite{1109.2604,1211.4873,1311.5896,1411.1335,1505.03867,1506.04149,1509.05323,1510.05378,1512.02472,1602.04788,1603.07387,1611.05048,1701.02737,1701.05869}. 
The stability of dark matter is guaranteed by a $Z_2$ symmetry. All the SM particles are $Z_2$-even.
There are a gauge singlet Weyl fermion and an SU(2) doublet Dirac fermion. They are $Z_2$-odd. 
The singlet, the doublet, and the Higgs boson form Yukawa interactions.
There is a CP phase in the dark sector, and thus the model naturally contains a pseudoscalar interaction.
The effect of this CP violation on the electron electric dipole moment (EDM) was discussed in Ref.~\cite{hep-ph/0510064,0705.4493}.
On the other hand, the CP violation effect on the DM phenomenology has hardly ever been discussed.\footnote{The analysis of DM phenomenology with CP phase 
for the light DM mass region, $m_{\text{DM}} \lesssim 100$~GeV, was discussed in~\cite{1411.1335}.}

In this paper, we examine the effect of the CP violation in the model. 
In particular, we focus on its effect on the cross section for the direct detection.
We show that the pseudoscalar interaction generated by the CP violation helps to evade the strong constraints 
from the direct detection experiments.
We also investigate the electron EDM with emphasis on the relation to the DM phenomenology.

The rest of this paper is organized as follows. In section~2, we briefly review the singlet-doublet dark matter model.
In section~3, we discuss the current constraints and prospects of this model from the viewpoint of DM direct searches and the electron EDM. 
We devote section~4 to our conclusion.

\section{Model}

In this section, we briefly review the model.
We introduce
a gauge singlet Majorana fermion ($\omega$)
and an SU(2)$_L$ doublet Dirac fermion with hypercharge $Y = 1/2$
which is composed of 
a left-handed Weyl fermion ($\eta = (\eta^{+}, \eta^0)^T$) 
and a right-handed Weyl fermion ($\xi^\dagger = ( (\xi^{-})^\dagger, \xi^{0 \dagger})^T$). 
We impose a $Z_2$ symmetry on the model. 
Under the $Z_2$ symmetry, all the SM particles are even, and all the fermions we introduced in the above are odd.
The lightest neutral $Z_2$-odd fermion is a DM candidate.

The mass and Yukawa interaction terms for the $Z_2$ odd particles are given by
\begin{align}
 {\cal L}_{int.} 
=&
- \frac{M_1}{2} \omega \omega
- M_2 e^{-i\phi} \xi \eta
- y \omega H^{\dagger} \eta
- y' \xi H \omega
+ (h.c.)
.
\end{align}
All the parameters have a CP violating phase in general, but we can eliminate three of them by
the redefinition of the fermion fields. 
We work in the basis where only the Dirac mass term has a phase,
and we explicitly write down the phase as $M_2 e^{-i \phi}$. 
In this basis, all the parameters except $\phi$ are positive.
After the Higgs field develops a vacuum expectation value (VEV), we find the following mass terms
\begin{align}
{\cal L}_{mass} &
=
- \frac{1}{2}
\left(
\begin{matrix}
 \omega & \eta^0 & \xi^0
\end{matrix}
\right)
\begin{pmatrix}
 M_1 & \frac{v}{\sqrt{2}} y & \frac{v}{\sqrt{2}} y' \\  
 \frac{v}{\sqrt{2}} y &  0 & M_2 e^{- i \phi} \\
 \frac{v}{\sqrt{2}} y'&  M_2 e^{- i \phi} &  0
\end{pmatrix}
\left(
\begin{matrix}
 \omega \\ \eta^0 \\ \xi^0
\end{matrix}
\right)
-
M_2 \xi^{-} \eta^{+}
+
(h.c.),
\label{eq:massMatrix}
\end{align}
where $v$ is the VEV of the Higgs boson, $v \simeq 246~\text{GeV}$.
We introduce $\lambda$, $\theta$, and $r$ for later convenience,
\begin{align}
 y = \lambda \sin\theta, \quad 
 y'= \lambda \cos\theta, \quad
 r = \frac{y}{y'} = \tan \theta.
\end{align}
The mass of the charged Dirac fermion is $M_2$.
After we diagonalize the mass matrix, we obtain three neutral Weyl fermions $(\chi^0_1, \chi^0_2, \chi^0_3)$
that are related to $(\omega, \eta^0, \xi^0)$ by a unitary matrix $V$ as follows.
\begin{align}
 \begin{pmatrix} \omega \\ \eta^0 \\ \xi^0  \end{pmatrix}
=
 \begin{pmatrix}
  V_{11} & V_{12} & V_{13} \\
  V_{21} & V_{22} & V_{23} \\
  V_{31} & V_{32} & V_{33}
 \end{pmatrix}
 \begin{pmatrix} \chi^0_1 \\ \chi^0_2 \\ \chi^0_3  \end{pmatrix}
,
\end{align}
where $\chi_1^0$ is the lightest neutral $Z_2$-odd field and thus is the DM candidate in this model.

It is sufficient to study $\phi$ in the range $0 \leq \phi \leq \pi$, although $-\pi < \phi \leq \pi$ in general.
Physics in the negative $\phi$ regime is related to physics in the positive $\phi$ regime
by the complex conjugate of the mass matrix, and thus of the mixing matrix.
If observables respect the CP symmetry, they do not depend on the sign of $\phi$.
All the processes for the relic abundance and the direct detection
are independent of the sign of $\phi$ because they respect the CP symmetry. 
While the CP violating EDM depends on the sign of $\phi$,
the sign of $\phi$ merely changes the sign of the EDM, 
so we focus only on the positive $\phi$ region.

The ratio of the two Yukawa couplings is important as we will see below. 
We denote it by $r = y/y'$ and focus on $0 < r \leq 1$.
Since both Yukawa couplings are positive, $r$ is positive definite.
For $r = 0$, we can eliminate $\phi$ by the redefinition of the fermion fields.
We do not discuss that situation because we aim to examine the CP violation effect in the dark sector. 
Physics for $1 < r < \infty$ is equivalent to physics for $0 < r <1$ 
by renaming $\eta^0$ and $\xi^0$ as $\xi^0$ and $\eta^0$ respectively as can be seen from the mass matrix.
Therefore it is sufficient to discuss for $0 < r \leq 1$.

\subsection{couplings}
The $Z_2$-odd fermions couple to the Higgs boson and the gauge bosons.
We can obtain the interaction terms by diagonalizing the mass matrix and going to the mass eigenbasis.
Four component notation is useful in calculations.
The mass eigenstates of the charged and neutral $Z_2$ odd particles in four component notation are given by
\begin{align}
 \Psi_+ = \begin{pmatrix} \eta^{+} \\ (\xi^{-})^{\dagger}\end{pmatrix}, \quad
 \Psi_j = \begin{pmatrix} \chi^0_{j} \\ \chi^{0}_{j}{}^{\dagger}\end{pmatrix}.
\end{align}
The interaction terms including $Z_2$-odd particles are\footnote{We have checked these interaction terms
are consistent with Ref.~\cite{1411.1335,1505.03867} up to conventions of the fields and the gauge couplings.}
\begin{align}
 {\cal L}_{int.}
\supset &
 - \sum_j \overline{\Psi_{+}} \gamma^{\mu} ( P_L c_{\chi_j}^L + P_R c_{\chi_j}^R) \Psi_j W^{+}_{\mu} 
\nonumber\\
&
 - \sum_j \overline{\Psi_{j}} \gamma^{\mu} ( P_L (c_{\chi_j}^L)^* + P_R (c_{\chi_j}^R)^* ) \Psi_{+} W^{-}_{\mu} 
\nonumber\\
&
- \left( \frac{e}{2 s_W c_W} (c_W^2 - s_W^2) Z_{\mu} + e A_{\mu} \right)  \overline{\Psi_{+}} \gamma^{\mu} \Psi_+ 
\nonumber\\
&
-\frac{1}{2} \sum_{j,k}  c_{Z \chi_j \chi_k} \overline{\Psi_{j}} \gamma^{\mu} \gamma^5\Psi_k Z_{\mu}
+\frac{1}{2} \sum_{j,k}  c_{Z \chi_j \chi_k}^p \overline{\Psi_{j}} i \gamma^{\mu} \Psi_k Z_{\mu}
\nonumber\\
&
-\frac{1}{2} \sum_{j,k} c_{h\chi_j \chi_k} \overline{\Psi_{j}} \Psi_k  h
+\frac{1}{2} \sum_{j,k} c_{h\chi_j \chi_k}^p \overline{\Psi_{j}} i \gamma_5 \Psi_k  h
,
\end{align}
where
\begin{align}
 c_{\chi_j}^L =& \frac{e}{\sqrt{2} s_W} V_{2j}, \\
 c_{\chi_j}^R =& \frac{e}{\sqrt{2} s_W} V_{3j}^{*}, \\
 c_{Z \chi_j \chi_k} &= \frac{e}{2 s_W c_W} \text{Re} (V_{2j}^{*} V_{2k} - V_{3j}^{*} V_{3k}), \label{eq:DM-Z-coup} \\
 c_{Z \chi_j \chi_k}^{p} &= \frac{e}{2 s_W c_W} \text{Im} (V_{2j}^{*} V_{2k} - V_{3j}^{*} V_{3k}),  \\
 c_{h\chi_j \chi_k}  &= \sqrt{2} \text{Re}(y V_{1j} V_{2k} + y' V_{1j} V_{3k}), \label{eq:DM-h-coup} \\
 c_{h\chi_j \chi_k}^p &= \sqrt{2} \text{Im}(y V_{1j} V_{2k} + y' V_{1j} V_{3k}).
\end{align}
Among these terms, the following interactions are particularly important for our discussion.
\begin{align}
- \frac{1}{2} c_{Z \chi_1 \chi_1} \bar{\Psi}_1 \gamma^{\mu} \gamma^5 \Psi_1 Z_{\mu} 
- \frac{1}{2} c_{h \chi_1 \chi_1} \bar{\Psi}_1 \Psi_1 h 
+ \frac{1}{2} c_{h \chi_1 \chi_1}^p \bar{\Psi}_1 i \gamma^5 \Psi_1 h 
.
\end{align}
All these three terms contribute to DM annihilation processes.
On the other hand, they play different roles in elastic scattering processes of the dark matter with nucleon.
The scalar interaction ($c_{h \chi_1 \chi_1}$) contributes to the spin-independent cross section ($\sigma_{\text{SI}}$), 
the $Z$ interaction ($c_{Z \chi_1 \chi_1}$) contributes to the spin-dependent cross section ($\sigma_{\text{SD}}$),
and the pseudoscalar interaction ($c_{h \chi_1 \chi_1}^p$) does not contribute to the scattering process due to the velocity
suppression.
Therefore, if the $c_{h \chi_1 \chi_1} = c_{Z \chi_1 \chi_1} = 0$,
then DM can completely evade the current strong constraints from the direct detection experiments.
Meanwhile, a nonzero value of $c_{h \chi \chi}^p$ can ensure the necessary annihilation cross section.

There are symmetries that may force $c_{h \chi_j \chi_k}^p$ and $c_{Z \chi_j \chi_k}$ to be zero.
In the CP invariant situations, $\phi = 0$ or $\pi$, then $c_{h \chi_j \chi_k}^p = 0$
because the pseudoscalar interaction originates from the CP violation.
If $y = y'$, then $c_{Z \chi_j \chi_k} = 0$ because the model becomes symmetric under the exchange of $\chi^0$ and $\xi^0$
as can be seen from Eq.~\eqref{eq:massMatrix}.
This symmetry implies $V_{2i} = V_{3i}$, and thus $c_{Z \chi_j \chi_k} = 0$ as can be seen from Eq.~\eqref{eq:DM-Z-coup}.

The scalar coupling $c_{h \chi_j \chi_k}$ can be zero as well. 
However, in contrast to $c_{h \chi_j \chi_k}^p$ and $c_{Z \chi_j \chi_k}$,
it becomes zero accidentally rather than by symmetries.
We discuss the condition on the parameters for vanishing $c_{h \chi_1 \chi_1}$ in the next subsection.

\subsection{Blind spot}

There is a condition that the scalar interaction of the DM vanishes, $c_{h \chi_1 \chi_1} = 0$.
The condition is called the blind spot~\cite{1109.2604,1211.4873,1311.5896}. 
In this subsection, we discuss the condition for the blind spot.

For the purpose of finding the condition that $c_{h\chi_1 \chi_1} = 0$, 
the expression of the scalar coupling given in Eq.~\eqref{eq:DM-h-coup} is not convenient.
We can also obtain the scalar coupling from the derivative of the dark matter mass
with respect to the VEV of the Higgs boson, $c_{h\chi_1 \chi_1} = \partial m_{\text{DM}}/\partial v$. 
The expression obtained in this way is useful to find the blind spot.
The DM mass satisfies the characteristic equation given by
\begin{align}
0 =&  
  \mDM^6
- \mDM^4 \left( M_1^2 + 2 M_2^2 + v^2 \lambda^2 \right)
\nonumber\\
&
+ \mDM^2 \left( 2 M_1^2 M_2^2 + \left(M_2^2 + \frac{v^2 \lambda^2}{2} \right)^2 - M_1 M_2 v^2 \lambda^2 \sin 2\theta \cos\phi \right)
\nonumber\\
&
- M_2^2 \left( M_1^2 M_2^2 - M_1 M_2 v^2 \lambda^2 \sin 2\theta \cos\phi + \frac{1}{4} v^4 \lambda^4 \sin^2 2\theta  \right)
,
\label{eq:eq_for_mDM}
\end{align}
where $\tan \theta = y/y' = r$. 
By differentiating this equation with respect to $v$ and setting $\partial m_{\text{DM}}/\partial v = 0$, 
we find
\begin{align}
0 =& 
  \mDM^4
- \mDM^2 \left( M_2^2 + \frac{v^2 \lambda^2}{2} - M_1 M_2 \sin 2\theta \cos\phi  \right)
- M_2^2 \sin 2\theta \left( M_1 M_2 \cos\phi - \frac{v^2}{2}\lambda^2 \sin 2\theta \right) 
.
\label{eq:eq_for_blind}
\end{align}
Using Eqs.~\eqref{eq:eq_for_mDM} and \eqref{eq:eq_for_blind}, we can obtain two relations.
For example, we can solve for $\mDM$ and $\lambda$,
\begin{align}
 \mDM
=&
\left(
 \cfrac{M_1^2 M_2^2 \sin^2 2\theta \sin^2\phi}{M_1^2 + M_2^2 \sin^2 2 \theta + 2 M_1 M_2 \sin2\theta \cos\phi}
\right)^{1/2}
,
\label{eq:blind_mDM}
\\
\lambda 
=&
\sqrt{
\frac{2(M_2^2 - \mDM^2)(\mDM^2 + M_1 M_2 \sin 2\theta \cos\phi)}{v^2 (M_2^2 \sin^2 2\theta - \mDM^2)}
}
.
\end{align}
This is the blind spot condition where $c_{h \chi_1 \chi_1} = 0$.
Eq.~\eqref{eq:blind_mDM} is given in Ref.~\cite{1411.1335}

For $\phi = 0$, there is no blind spot because it requires $\mDM = 0$.
For $\phi = \pi$, $\mDM$ is non-zero if the denominator of Eq.~\eqref{eq:blind_mDM} is zero
and we find the following blind spot condition for $\phi = \pi$,
\begin{align}
 m_{DM} =& M_1 = M_2 \sin 2\theta. \label{eq:blind_NOCPV}
\end{align}
This is the same as the blind spot condition given in Refs.~\cite{1311.5896, 1505.03867}.\footnote{
Note that 
$m_{\text{DM}} = M_2 \sin 2\theta = - M_2 \cos \pi \sin 2\theta $,
and 
$\sin 2\theta$ in Refs.~\cite{1311.5896, 1505.03867} corresponds to $\cos\phi \sin 2\theta$ in our notation.
}

\subsection{EDM}
\begin{figure}[tb]
\centering
\includegraphics[width=0.3\hsize]{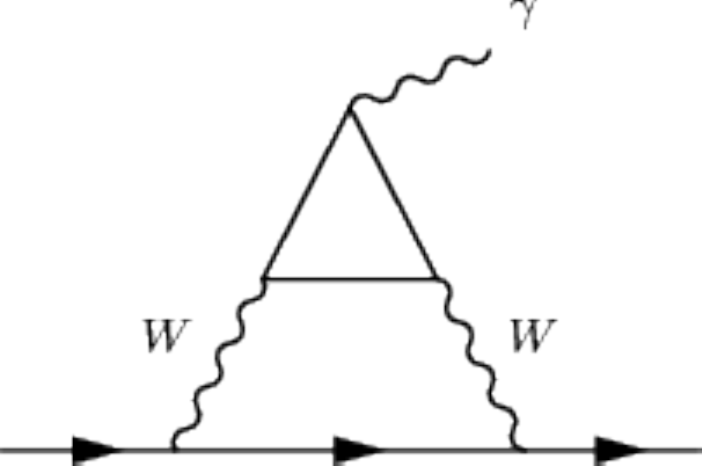} \label{fig:wwgamma}
\caption{Barr-Zee type contributions to the EDM. The $Z_2$-odd charged and neutral fermions run in the triangle part.
}
\label{fig:BZ-EDM}
\end{figure}

This model predicts a new contribution to the EDM because of the new source of CP violation $\phi$.
The electron EDM is an important complement to direct detection experiments as we will see in Sec.~\ref{sec:results}. 
We summarize the formulae here.
The leading contribution comes from the Barr-Zee type diagram shown in Fig.~\ref{fig:BZ-EDM}.
The electron EDM is defined through
\begin{align}
 {\cal H}_{eff}
=&
 i \frac{d_e}{2}
\bar{\psi}_{e}
\sigma_{\mu \nu} \gamma_5
\psi_{e}
F^{\mu \nu}
.
\end{align}
We find~\cite{1411.1335}.
\begin{align}
 \frac{d_e}{e}
=&
- 
\frac{2 \alpha}{(4\pi)^3 s_W^2} \sqrt{2} G_F
m_{\chi^{\pm}} m_e
\sum_{j=1}^3
\text{Im}(V_{2j} V_{3j})
m_{\chi^0_j}
{\cal I}_j
,
\end{align}
where
\begin{align}
{\cal I}_j
=&
 \int_0^1 dz
\frac{1-z}{m_{\chi^{\pm}}^2 (1-z) + m_{\chi^0_j}^2 z - m_W^2 z (1-z)}
\ln \frac{ m_{\chi^{\pm}}^2 (1-z) + m_{\chi^0_j}^2 z}{m_W^2 z (1-z)}
.
\end{align}

\section{Current and future prospects on direct detection and EDM}
\label{sec:results}
In this section, we discuss the current constraints from the dark matter direct searches~\cite{1608.07648, 1611.06553}
and prospects for the direct searches~\cite{1310.8327}.
We assume the thermal relic scenario. 
We have calculated the relic abundance and the scattering cross section of dark matter with nucleon
by using \texttt{micrOMEGAs v4.3.1}~\cite{1407.6129}.
We choose the value of $\lambda$ so as to reproduce the measured dark matter density, $\Omega h^2= 0.1198 \pm 0.0015$~\cite{1502.01589}.

\begin{figure}[tb]
\includegraphics[width=0.32\hsize]{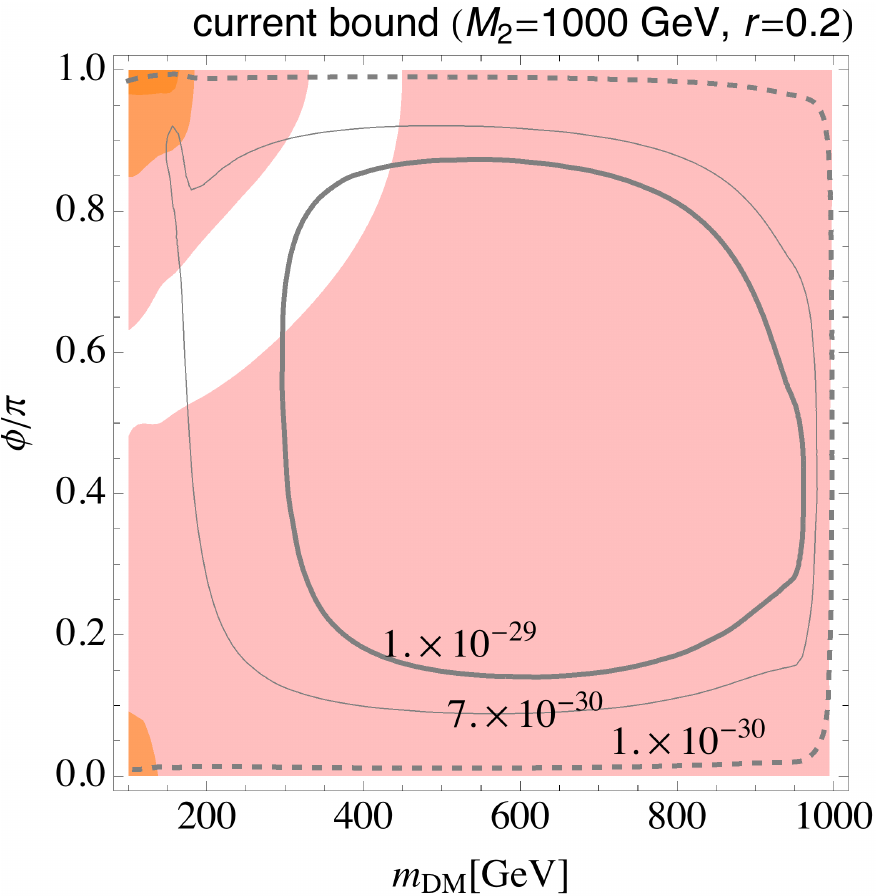}
\includegraphics[width=0.32\hsize]{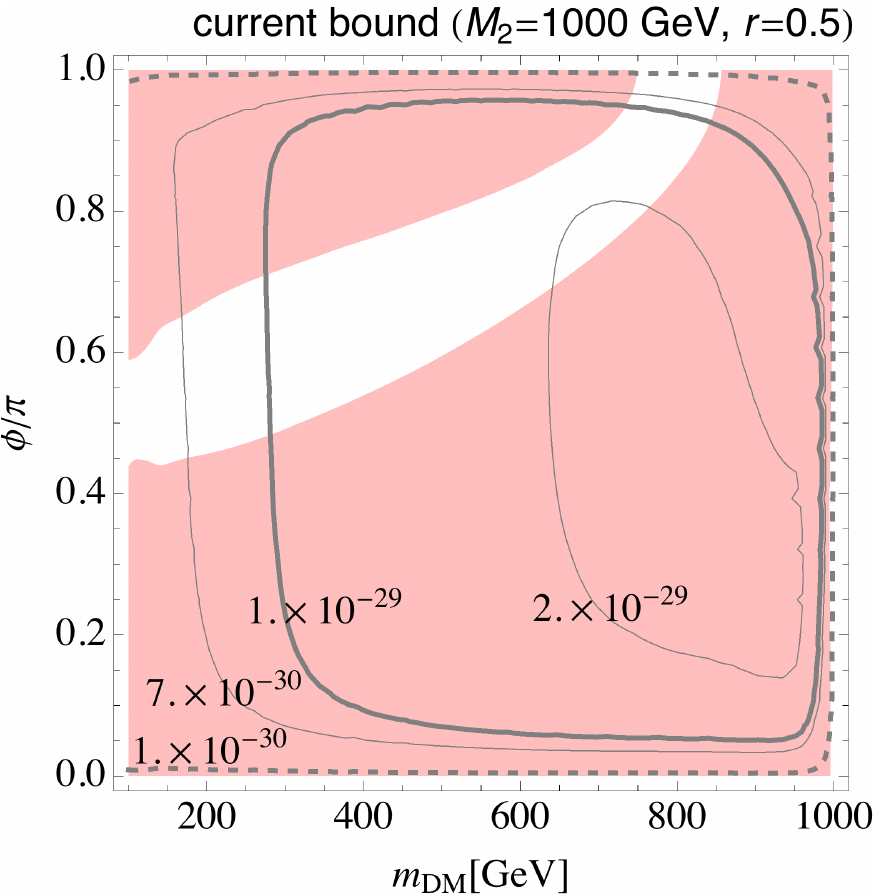}
\includegraphics[width=0.32\hsize]{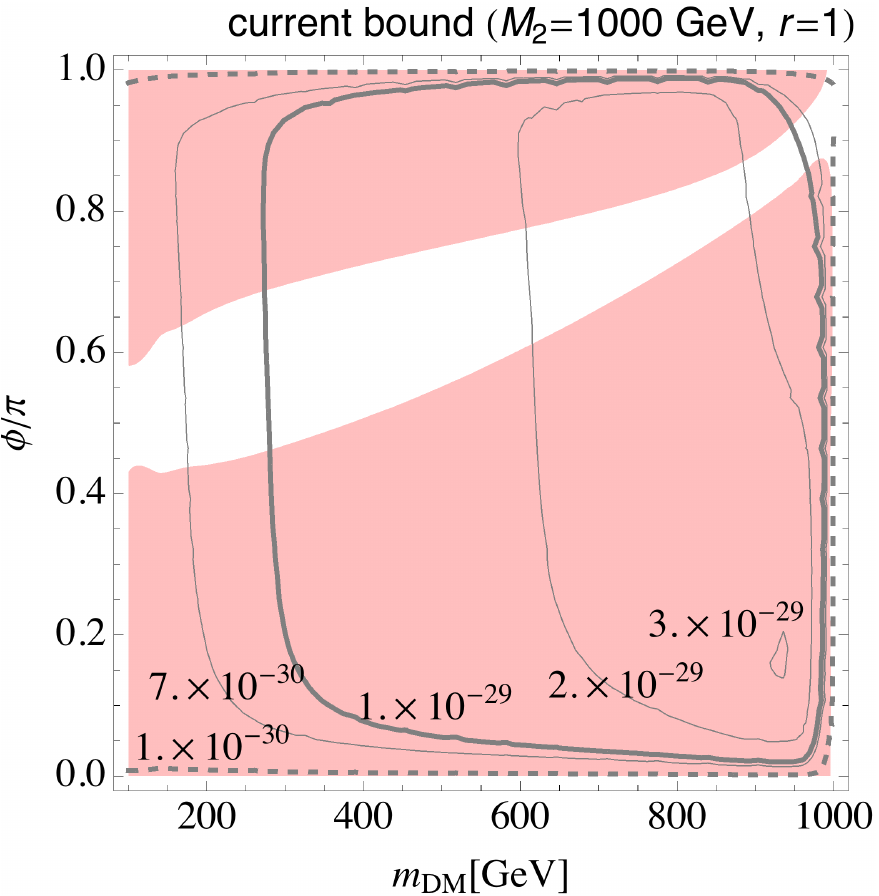}
\includegraphics[width=0.32\hsize]{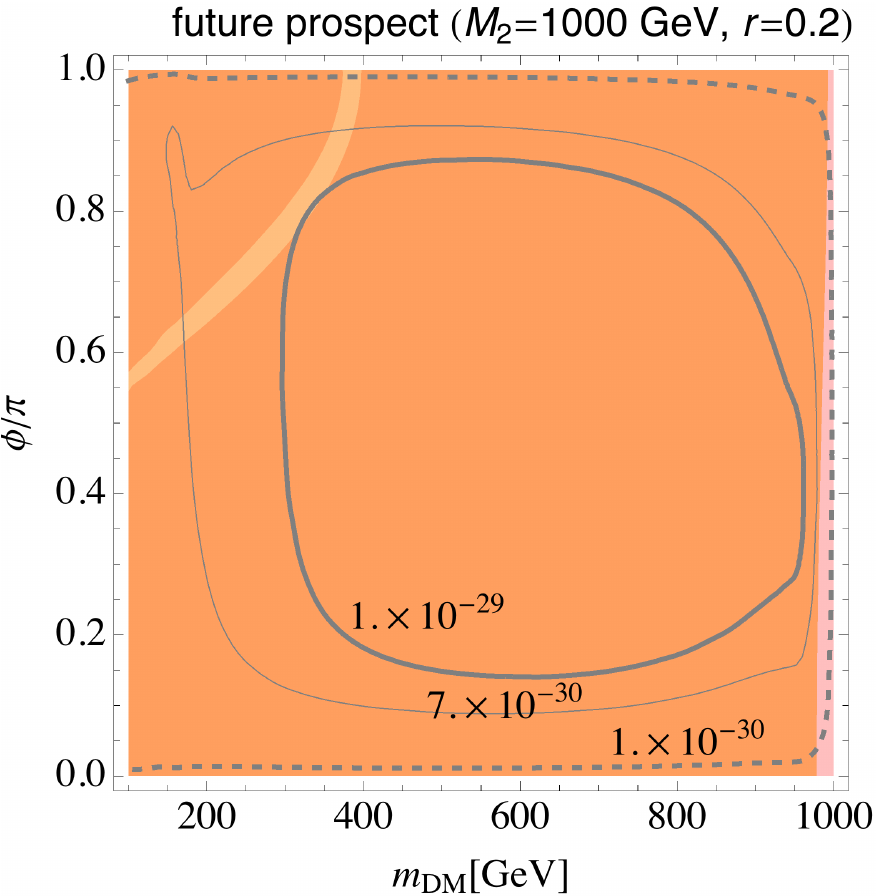}
\includegraphics[width=0.32\hsize]{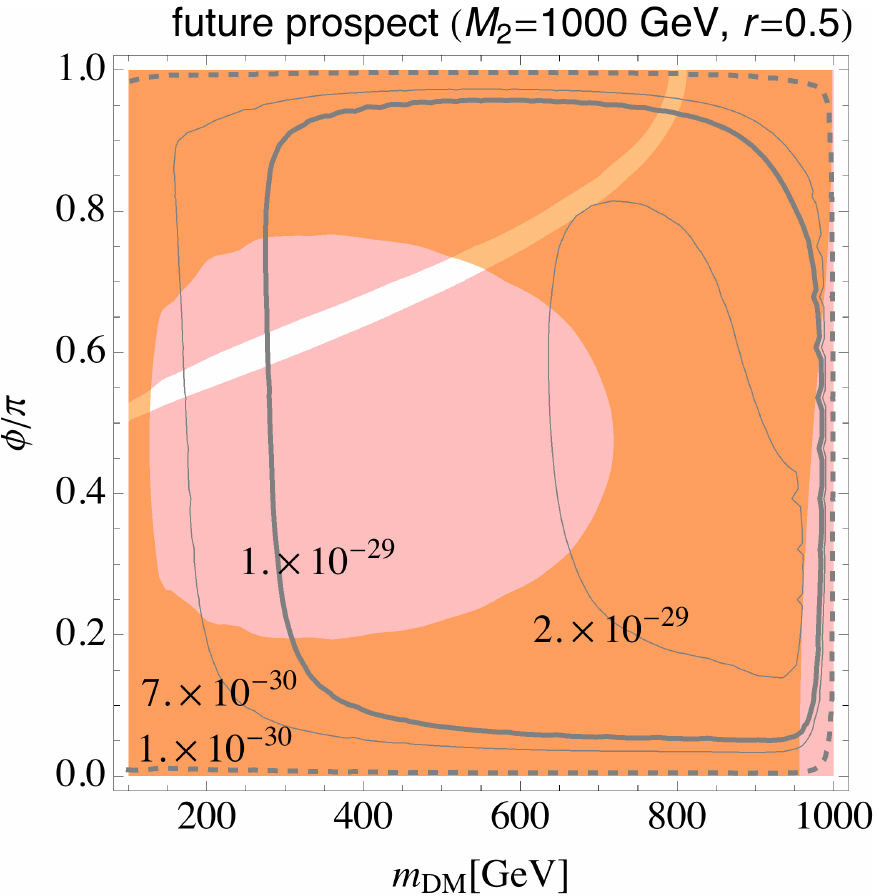}
\includegraphics[width=0.32\hsize]{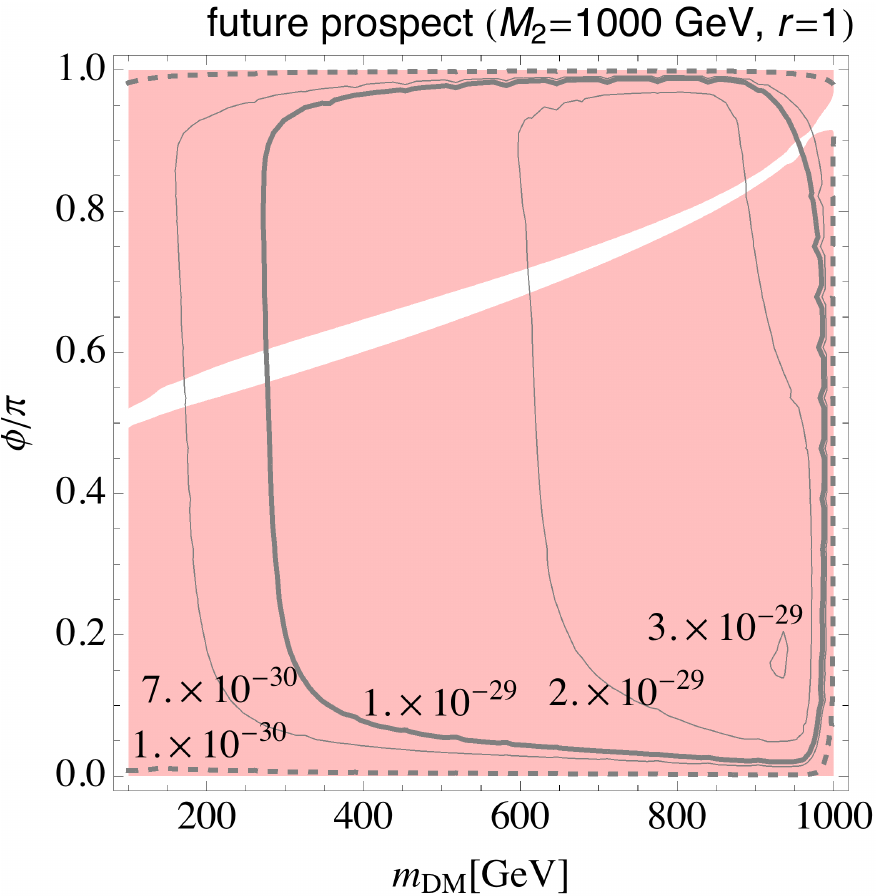}
\caption{
The current bound and prospects.
The red regions in the upper panels are excluded by the constraint on the $\sigma_{\text{SI}}$ from LUX experiment~\cite{1608.07648}.
The orange regions in the upper panels are excluded by the constraint on $\sigma_{\text{SD}}$ from PandaX-II experiment~\cite{1611.06553}.
The red and orange regions in the lower panels show the prospects for $\sigma_{\text{SI}}$ and $\sigma_{\text{SD}}$, respectively.
We use the projection for LZ experiment in Ref.~\cite{1310.8327} for the future prospects.
The contour lines show the absolute values of the model prediction of the electron EDM.
}
\label{fig:M2_1000}
\end{figure}

The current constraints from the dark matter direct searches and future prospects are shown in Fig.~\ref{fig:M2_1000}.
Here we take $M_2 = 1000$~GeV. 
The red regions are excluded by the constraint on $\sigma_{\text{SI}}$ from LUX experiment~\cite{1608.07648},
and the orange regions are excluded by the constraint on $\sigma_{\text{SD}}$ from PandaX-II experiment~\cite{1611.06553}.
We use the projection for LZ experiment in Ref.~\cite{1310.8327} for the future prospects of $\sigma_{\text{SI}}$ and $\sigma_{\text{SD}}$.
We find that the constraint on $\sigma_{\text{SI}}$ already excludes a large region of the parameter space.
However, the spin-independent scattering process cannot cover some of the regions due to the existence of the blind spot.
In those regions, the spin-dependent scattering process is helpful.
The constraint on $\sigma_{\text{SD}}$ is currently much weaker than the constraint from $\sigma_{\text{SI}}$.
However, it plays an important role in future as we can see from the lower panels in the figure.
The spin-dependent scattering process can cover the parameter space where the spin-independent scattering process cannot.
In $r \sim 1$ regime, $\sigma_{\text{SD}}$ becomes small 
because it depends on $c_{Z \chi_1 \chi_1}$ that becomes zero for $r = 1$ as we discussed in the previous 
section.

We also show the absolute value of the electron EDM by the contours in the figure.
The current bound is $|d_e| \leq 8.7 \times 10^{-29}$~e cm by the ACME experiment~\cite{1310.7534}.
We find $|d_e| \lesssim 3 \times 10^{-29}$~e cm in the figure.
Therefore there is no constraint from the current upper bound on the electron EDM.
We can expect the regions where $|d_e| \gtrsim {\cal O}(10^{-30})$~e cm is detectable in future~\cite{Sakemi:2011zz,1208.4507,Kawall:2011zz,1502.04317}.
This model predicts $|d_e| \gtrsim 10^{-30}$~e cm in nearly all of the parameter space,
and thus the electron EDM is expected to be probed in future.

\begin{figure}[tb]
\includegraphics[width=0.32\hsize]{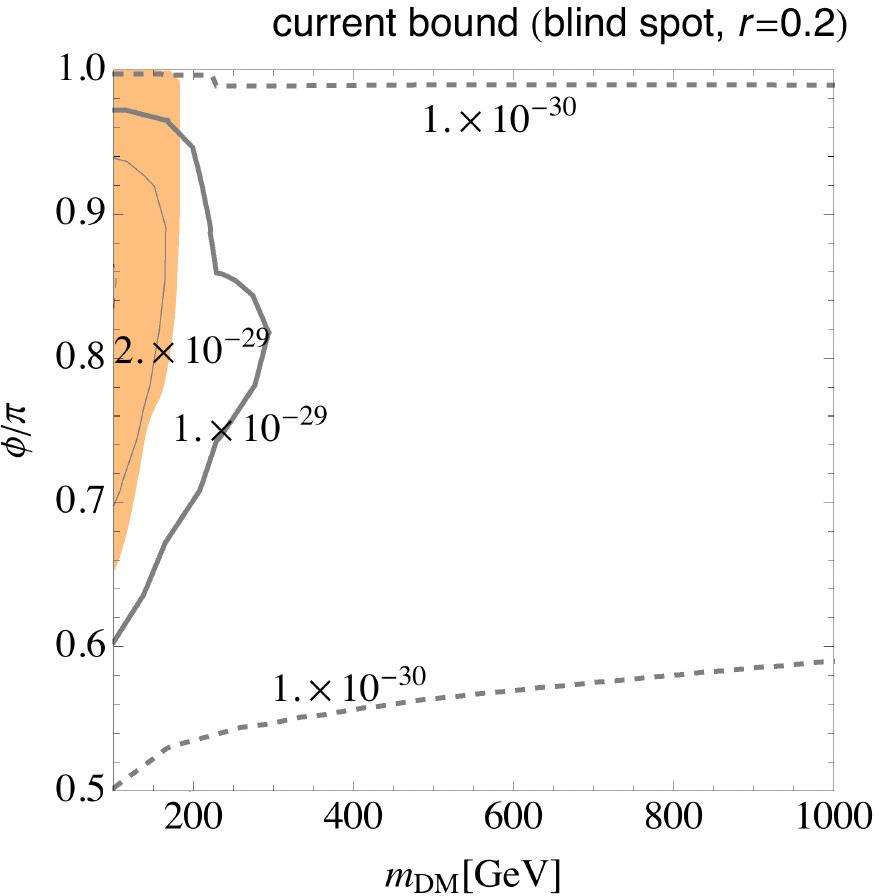}
\includegraphics[width=0.32\hsize]{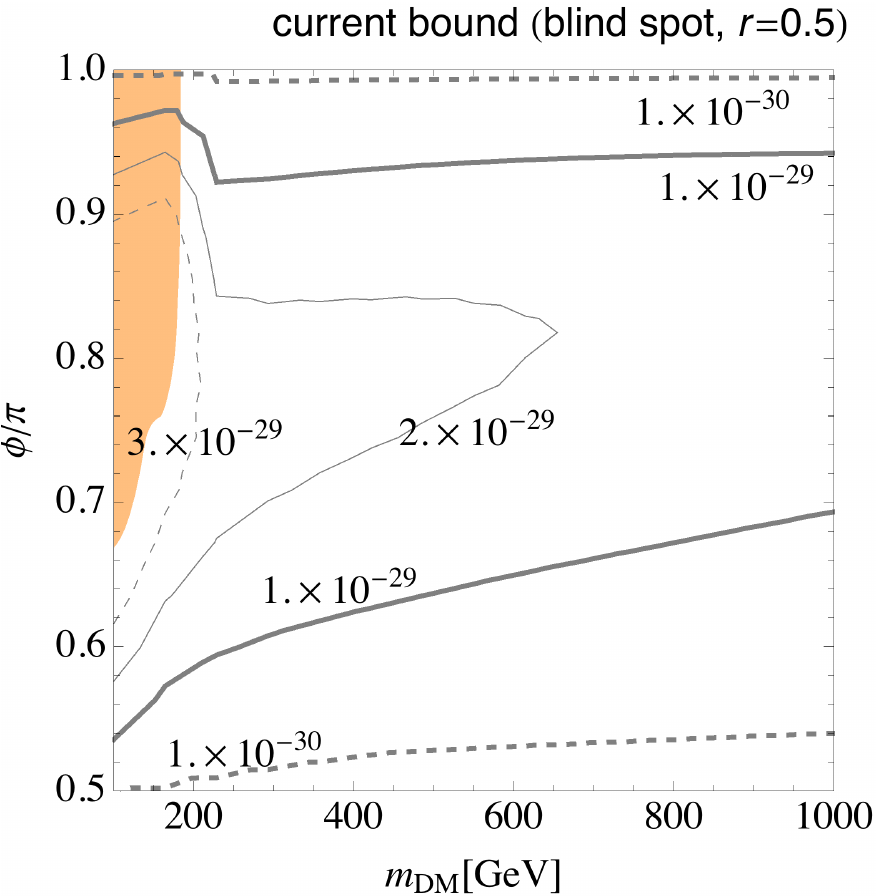}
\includegraphics[width=0.32\hsize]{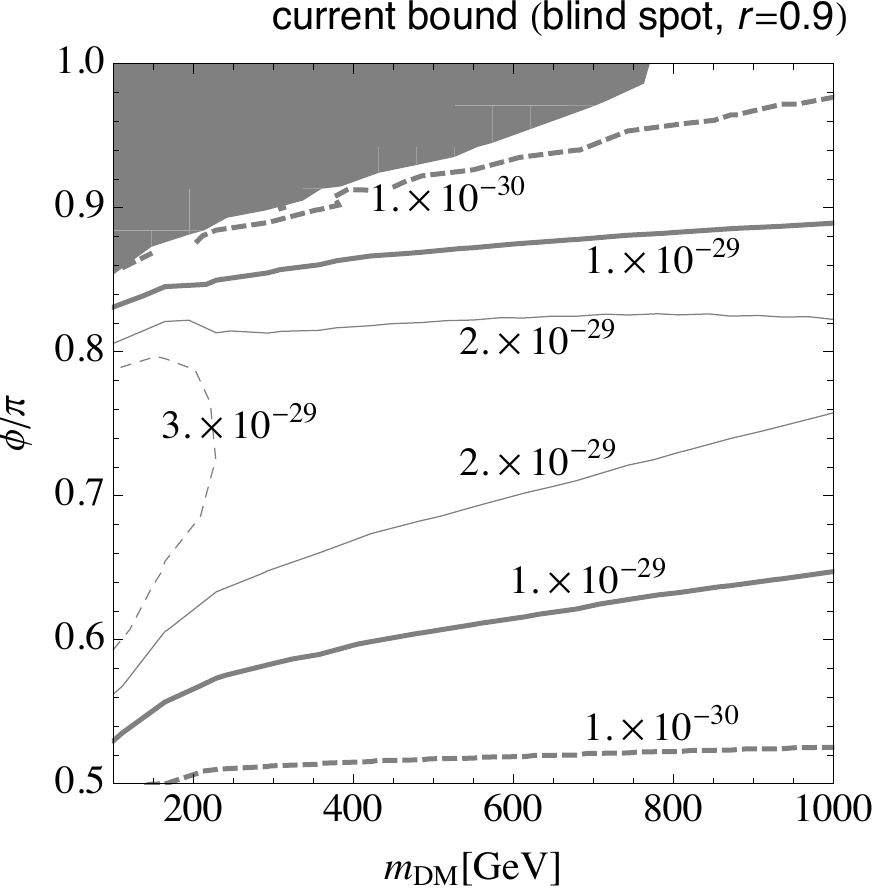}
\includegraphics[width=0.32\hsize]{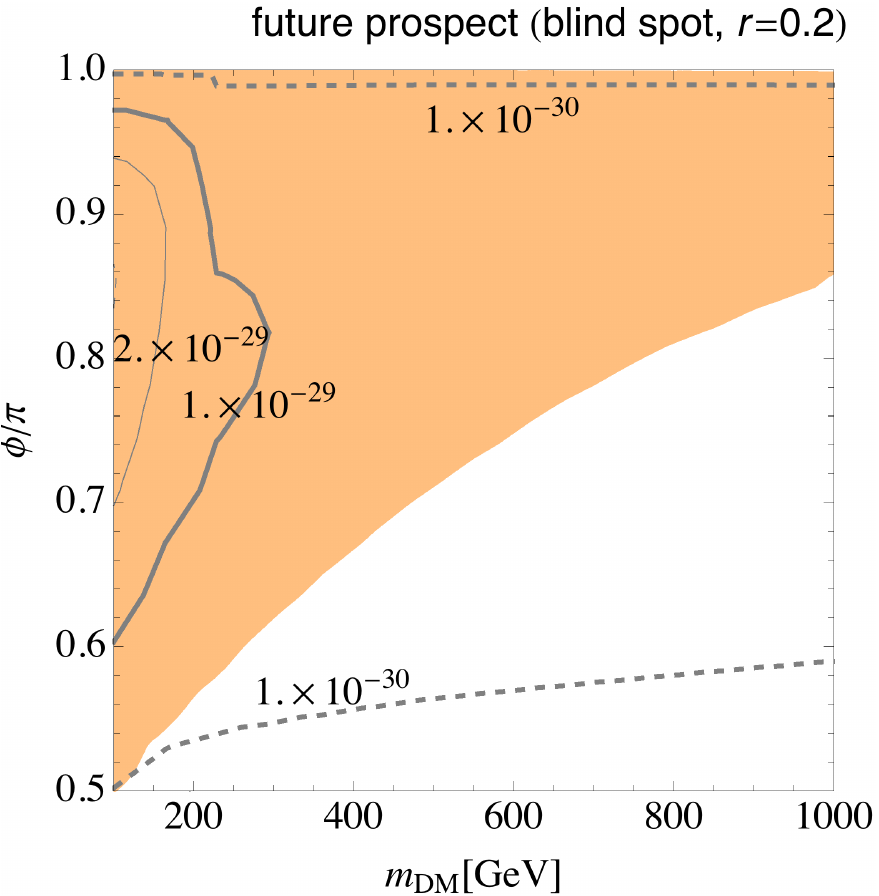}
\includegraphics[width=0.32\hsize]{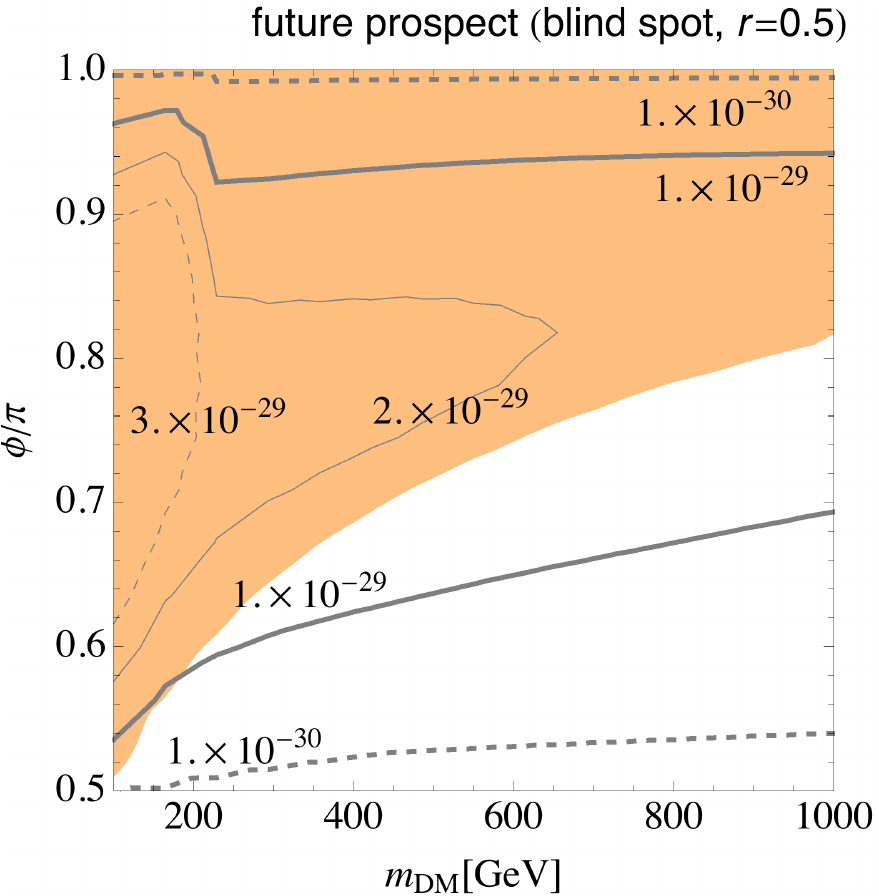}
\includegraphics[width=0.32\hsize]{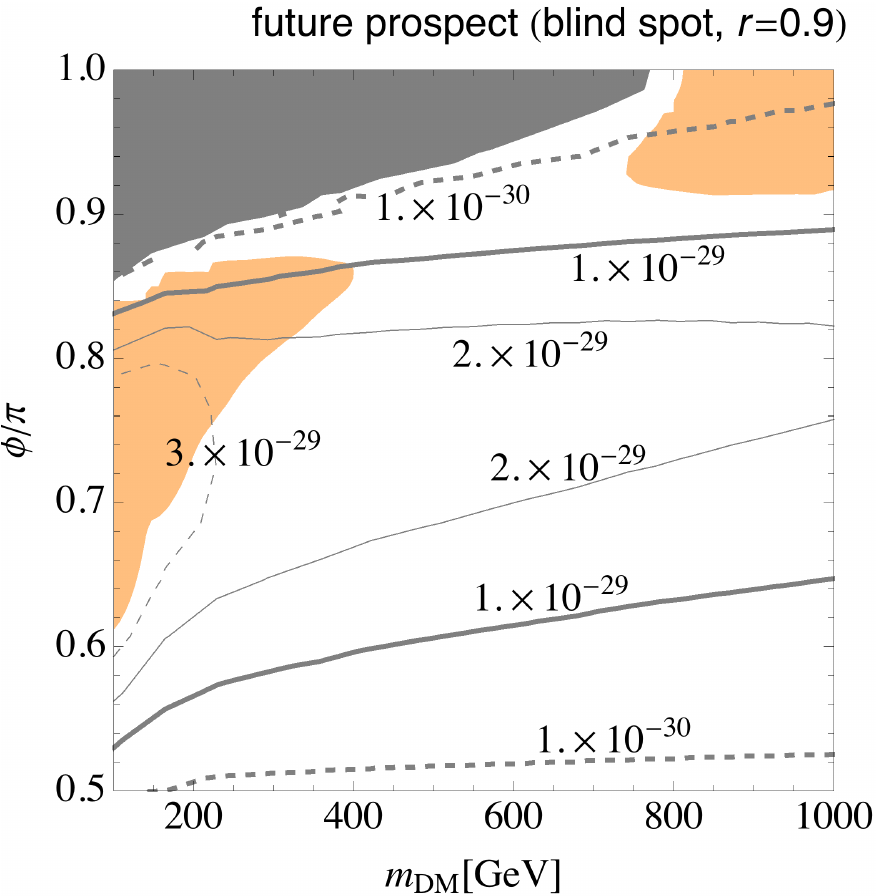}
\caption{
The current constraints and prospects in the blind spot.
In the gray regions, we cannot obtain the DM thermal relic abundance that matched the measured value of DM density. 
The notations of the other colors and the contours are the same as in Fig.~\ref{fig:M2_1000}.
}
\label{fig:blind_case}
\end{figure}

We next focus on the blind spot where $\sigma_{\text{SI}} = 0$. 
Here we determine the value of $M_2$ so that the blind spot condition is satisfied.
We show the current bound and prospects of $\sigma_{\text{SD}}$ with the electron EDM in Fig.~\ref{fig:blind_case}.
In the gray region and for $0 \leq \phi \lesssim \pi/2$, 
the DM thermal relic abundance does not match the measured value of DM density
while satisfying the blind spot condtion.
We find that the spin-dependent scattering process is a powerful tool to detecting DM in future. 
Although some regions of the parameter space cannot be covered by the spin-dependent scattering process,
the electron EDM is within the reach of the future experiments in the large parts of those regions.
Therefore the electron EDM is an important complement to the direct detection in the blind spot.

\begin{figure}[tb]
\includegraphics[width=0.32\hsize]{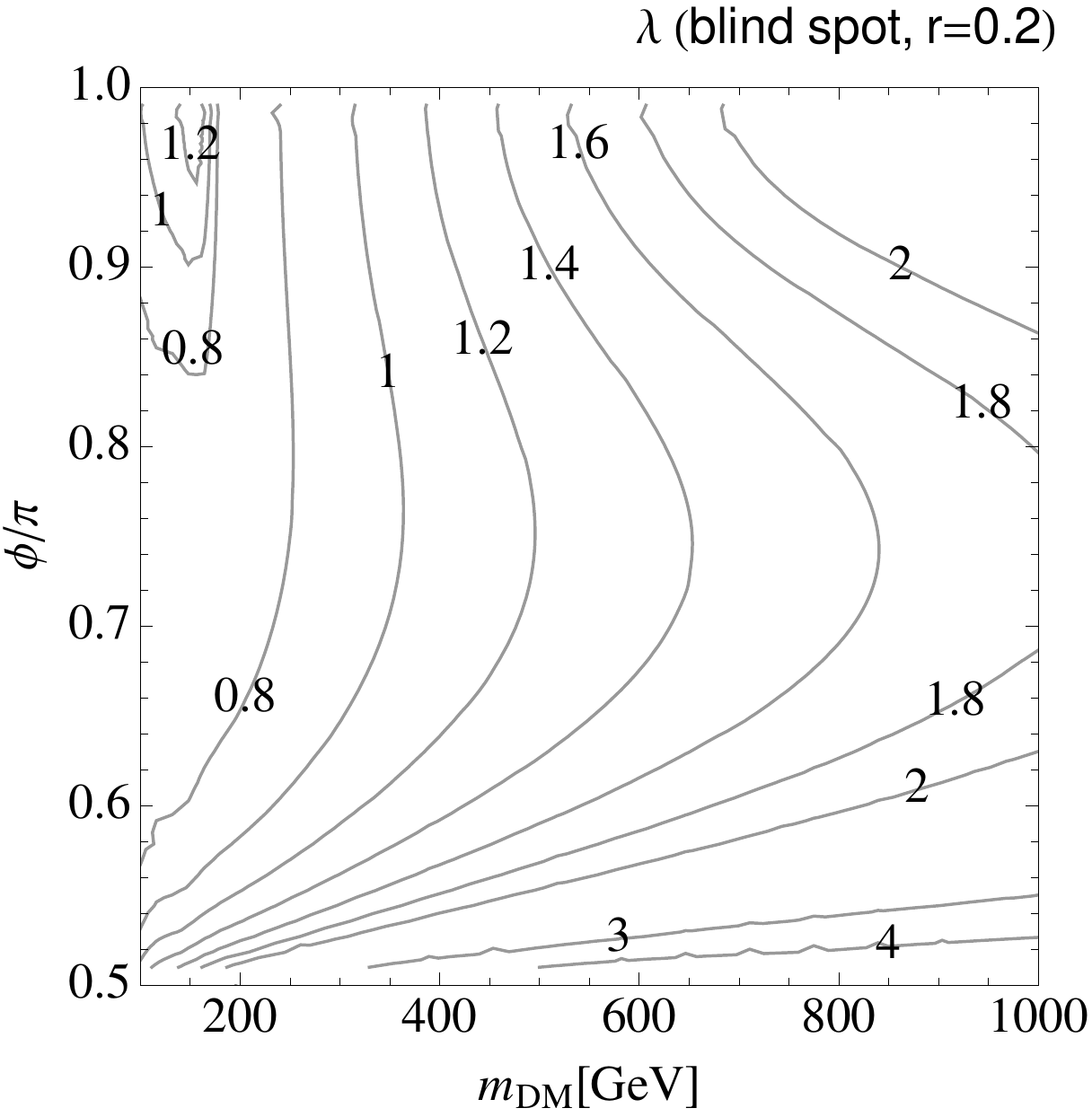}
\includegraphics[width=0.32\hsize]{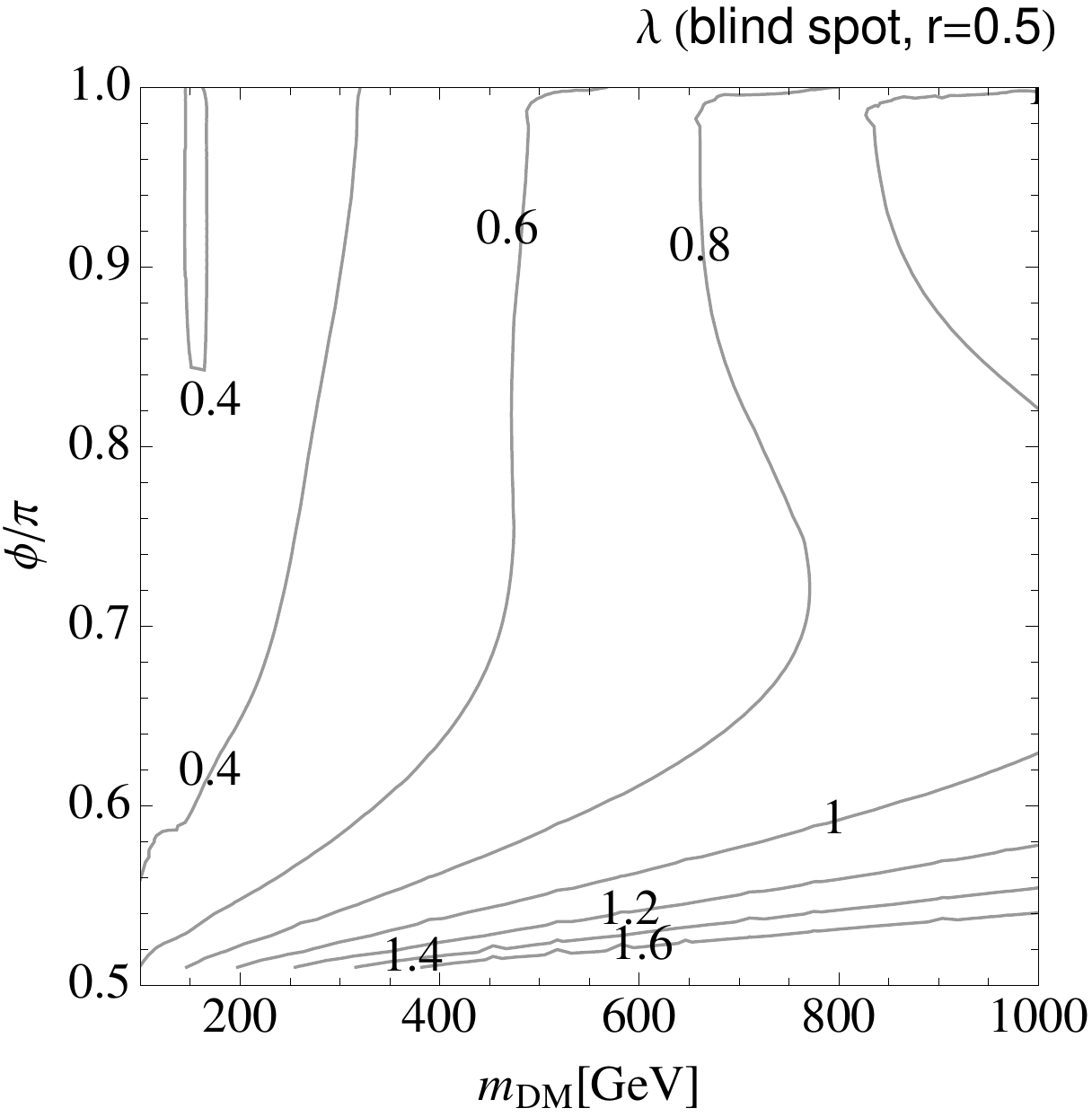}
\includegraphics[width=0.32\hsize]{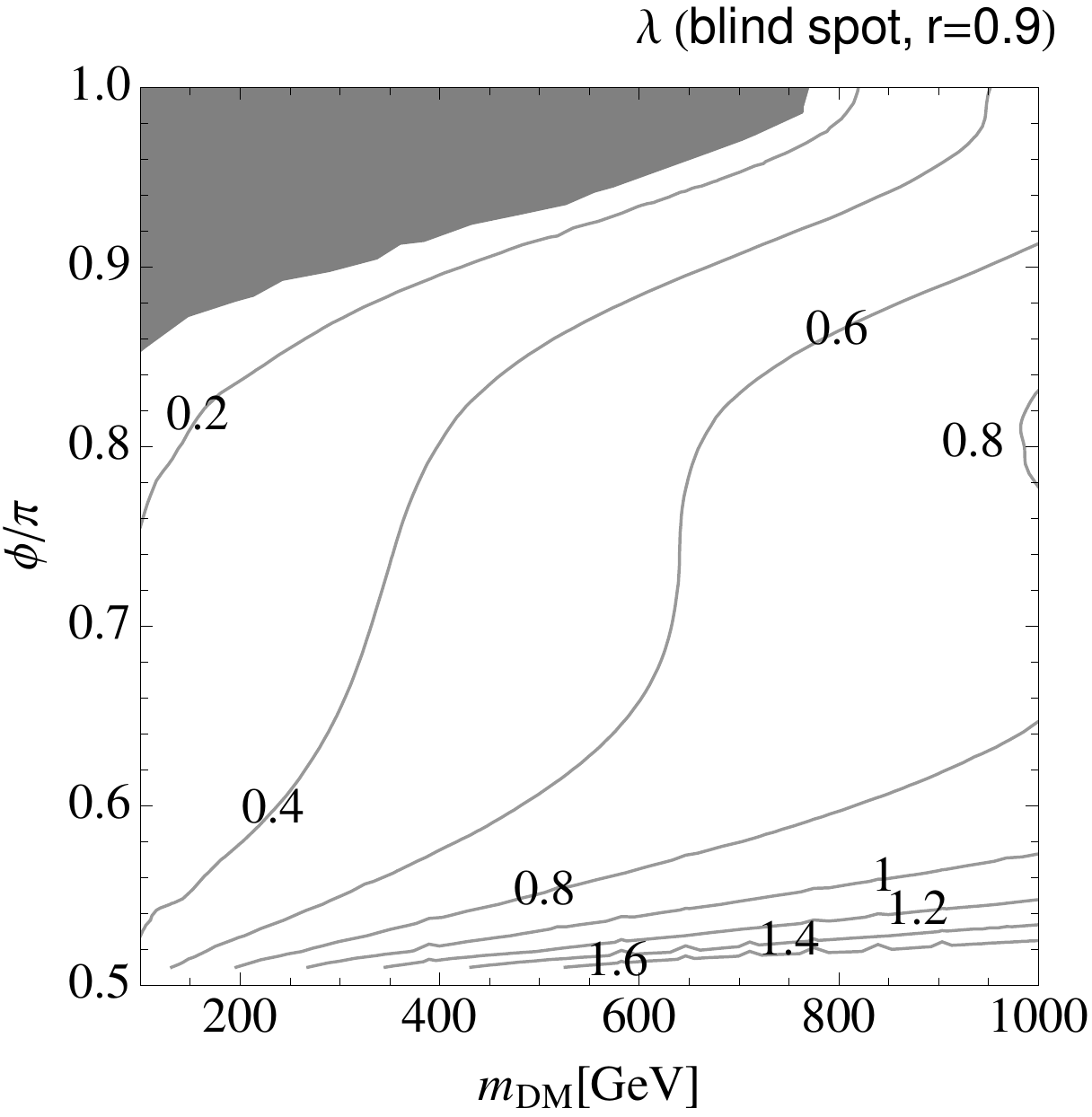}
\includegraphics[width=0.32\hsize]{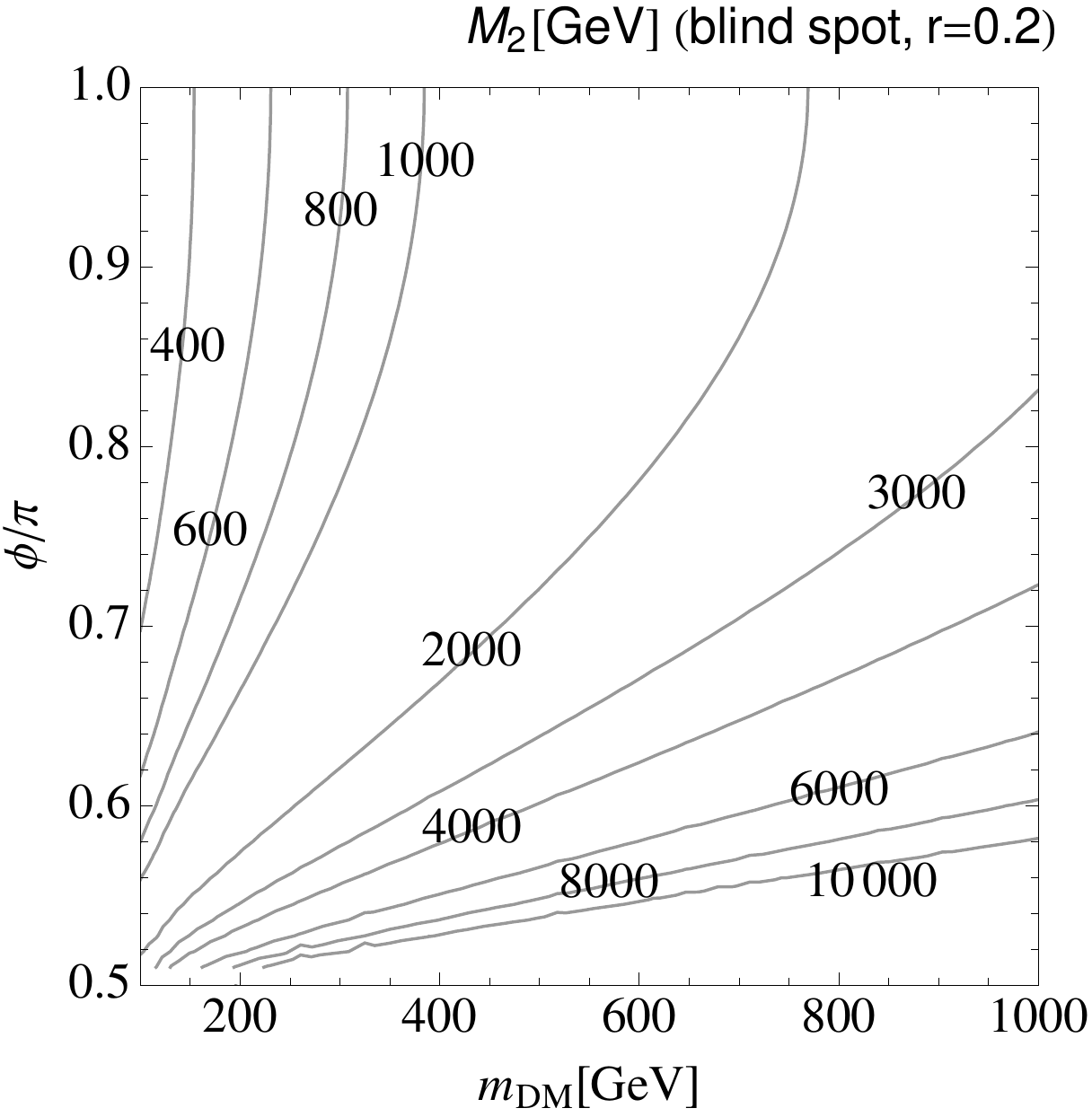}
\includegraphics[width=0.32\hsize]{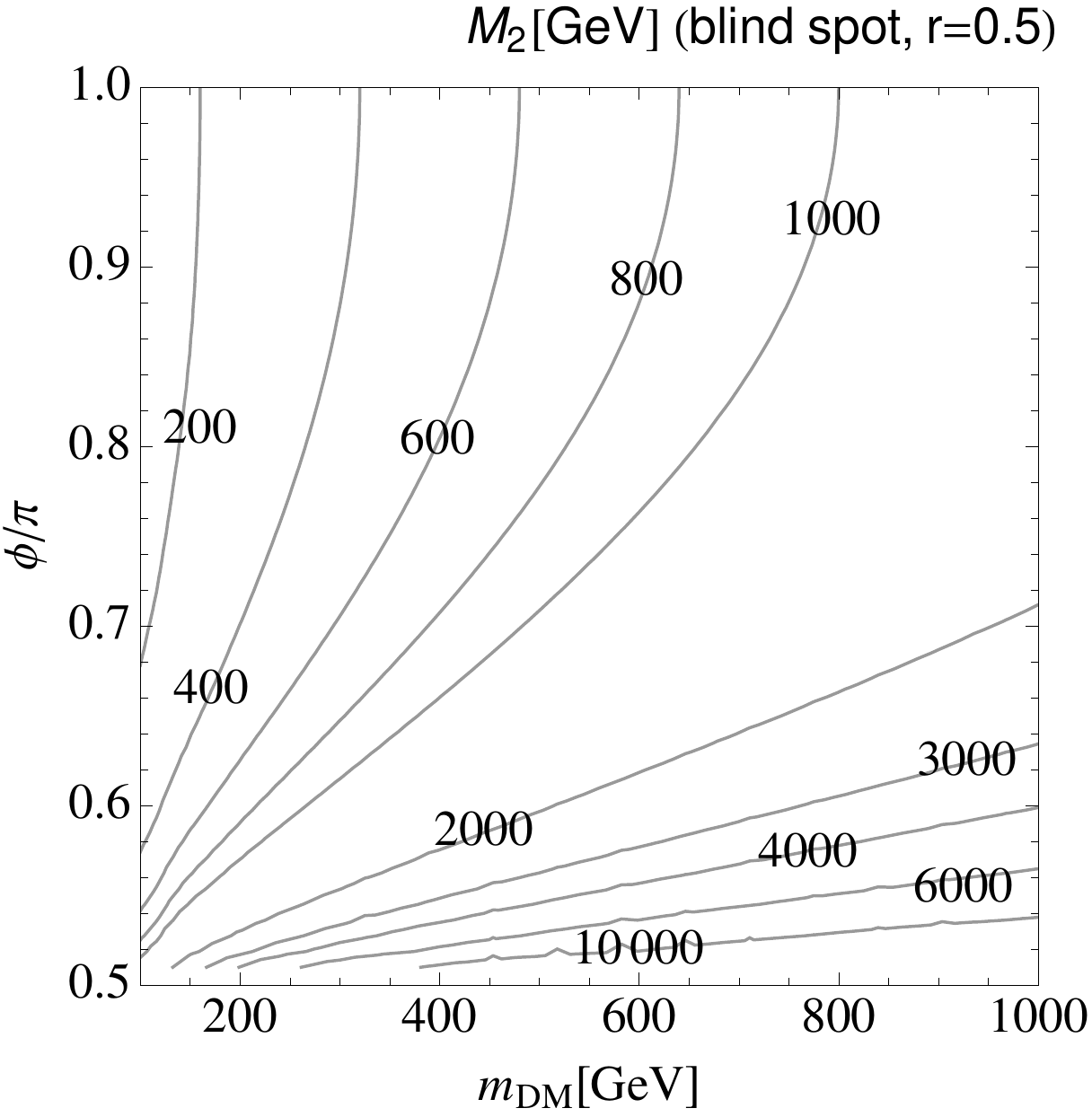}
\includegraphics[width=0.32\hsize]{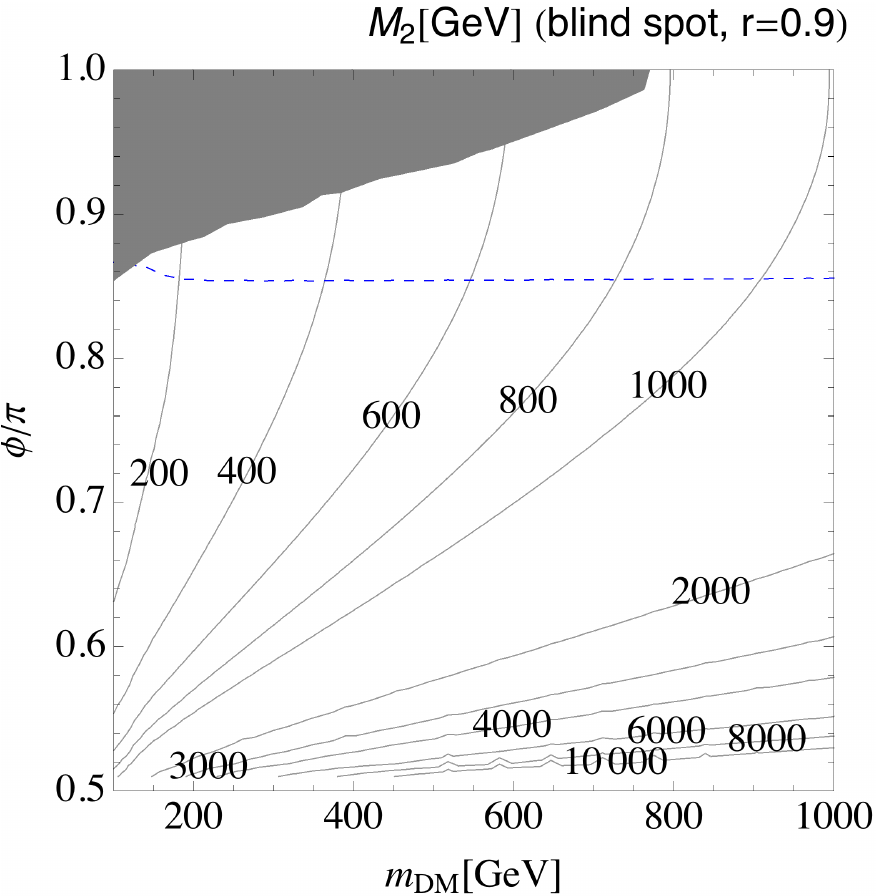}
\caption{
The values of $\lambda$ and $M_2$ in the blind spot.
The mass of the charged $Z_2$-odd particle is $M_2$.
The other masses, $m_{\chi_1}$ and $m_{\chi_2}$, are almost same as $M_2$.
In the region above the blue dashed curve, $M_2 - m_{\text{DM}} < 0.1 m_{\text{DM}}$,
and co-annihilation processes for the relic abundance are efficient and cannot be ignored. 
}
\label{fig:lam_and_M2}
\end{figure}

Figure~\ref{fig:lam_and_M2} shows that the values of $\lambda$ and $M_2$ in the blind spot.
We can see that $\lambda \lesssim 1$ in the wide regions of the parameter space.
The mass of the charged $Z_2$-odd particle is $M_2$,
and the masses of heavier $Z_2$-odd neutral particles are almost same as $M_2$ in Fig.~\ref{fig:blind_case} and \ref{fig:lam_and_M2}.
Therefore the lower panels in Fig.~\ref{fig:lam_and_M2} shows that the masses of the $Z_2$-odd particles other than the dark matter candidate.
In the region above the blue dashed curve in the lower-right panel, 
the mass differences between the dark matter and the other $Z_2$-odd particles are within the 10\% of the dark matter mass,
$M_2 - m_{\text{DM}} < 0.1 m_{\text{DM}}$. 
In that region, we find that co-annihilation processes for the relic abundance are efficient and cannot be ignored.
For $r$ = 0.1 and 0.5, we have checked the co-annihilation processes are negligible.
We also have checked the constraint from the $ST$ parameters~\cite{PRLTA.65.964,PHRVA.D46.381},
and found that the model is consistent with the current electroweak precision measurements~\cite{Olive:2016xmw}
in all the regions of the parameter space shown here.

\begin{figure}[tb]
\centering
\includegraphics[width=0.32\hsize]{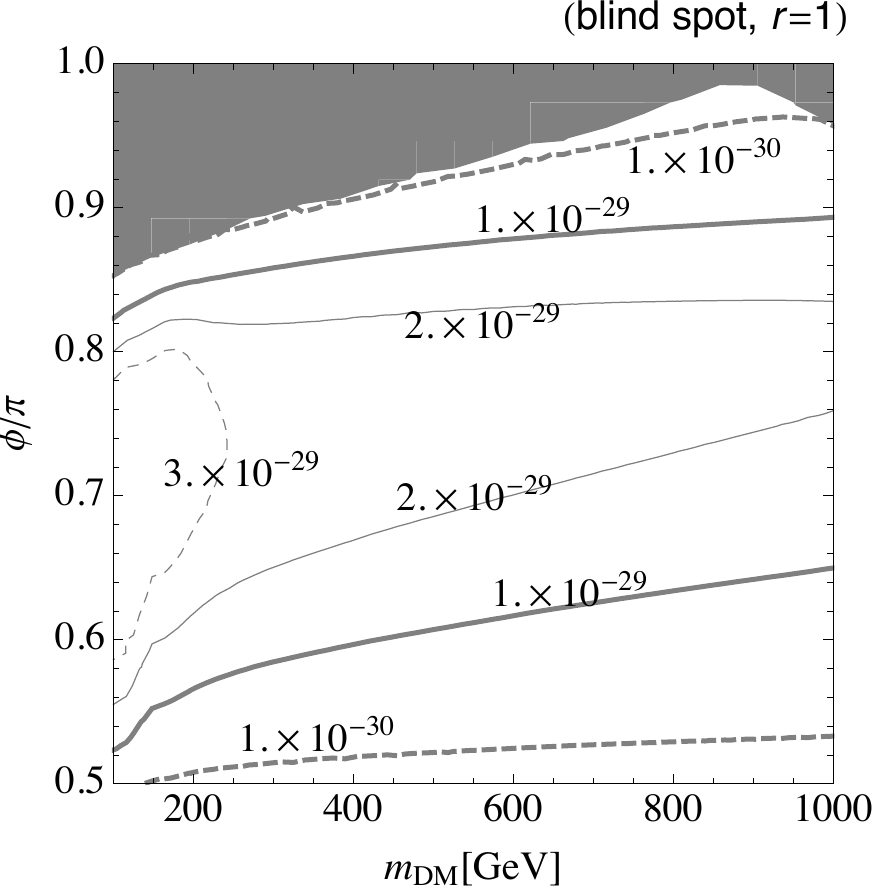}
\caption{
The contours show the electron EDM in a special case where dark matter completely evade the direct detection via scattering with nucleon,
namely $\sigma_{\text{SI}} = \sigma_{\text{SD}} = 0$.
In the gray regions, we cannot obtain the DM thermal relic abundance that matched the measured value of DM density. 
}
\label{fig:complete_blind_case}
\end{figure}

Finally, we discuss the blind spot with $r = 1$. 
This is a special parameter choice where the model can completely evade the constraints from the direct detection, namely
$\sigma_{\text{SI}} = \sigma_{\text{SD}} = 0$.
In this case, both $c_{h \chi_1 \chi_1}$ and $c_{Z \chi_1 \chi_1}$ vanish,
and thus we have to rely on the pseudoscalar coupling $c_{h \chi \chi}^p$ to realize the thermal relic scenario.
In the white region in Fig.~\ref{fig:complete_blind_case}, we can obtain the DM thermal relic abundance that matches the measured value of DM density.
Therefore this model is viable even if direct detection experiments do not find any signal.
We also show the absolute value of the electron EDM in the figure. 
The value is within the reach of the future experiments in most of the parameter space,
and thus we can confirm the validity of this model by the observation of the electron EDM.

\section{Conclusion}

The recent progress of the dark matter direct detection experiments gives stringent constraint on the
scattering cross section of dark matter with nucleon. 
Dark matter models in the thermal relic scenario have to evade this constraint. 
One simple way to evade the constraint is to rely on pseudoscalar interactions of a fermionic dark matter candidate.

We have studied the singlet-doublet dark matter model with a special emphasis on the CP violation effect.
The model contains pseudoscalar interactions that originate from the CP violation in the dark sector.
Even if dark matter direct detection experiments do not observe any signal,
the model can explain the dark matter thermal relic abundance that matches the measured value of DM density thanks to the CP violation in the dark sector.
The CP violation in the dark sector induces the electron EDM. 
We have shown that its value is larger than $10^{-30}$~e cm in a large region of the parameter space where we can obtain the measured value of DM density.
This value is within the reach of the future experiments, 
and the electron EDM is thus strongly expected to be observed.
Therefore, the electron EDM measurement is an important complement to the direct dark matter detection experiments for testing this model.

\end{document}